# Adaptive Distributed Hierarchical Sensing Algorithm for Reduction of Wireless Sensor Network Cluster-Heads Energy Consumption

Gal Oren
Department of Computer Science,
Ben-Gurion University of the Negev,
POB 653, Be'er-Sheva, Israel
Department of Physics,
Nuclear Research Center-Negev,
P.O.B. 9001, Be'er-Sheva, Israel
orenw@post.bgu.ac.il

Leonid Barenboim
Department of Mathematics and Computer Science,
The Open University of Israel,
P.O.B. 808, Ra'anana, Israel
leonidb@openu.ac.il

Harel Levin
Department of Mathematics and Computer Science,
The Open University of Israel,
P.O.B. 808, Ra'anana, Israel
Department of Physics,
Nuclear Research Center-Negev,
P.O.B. 9001, Be'er-Sheva, Israel
harellevin@gmail.com

*Abstract*—**Energy efficiency is a crucial performance metric in sensor networks, directly determining the network lifetime. Consequently, a key factor in WSN is to improve overall energy efficiency to extend the network lifetime. Although many algorithms have been presented to optimize the energy factor, energy efficiency is still one of the major problems of WSNs, especially when there is a need to sample an area with different types of loads. Unlike other energy-efficient schemes for hierarchical sampling, our hypothesis is that it is achievable, in terms of prolonging the network lifetime, to adaptively re-modify CHs sensing rates (the processing and transmitting stages in particular) in some specific regions that are triggered significantly less than other regions. In order to do so we introduce the Adaptive Distributed Hierarchical Sensing (ADHS) algorithm. This algorithm employs a homogenous sensor network in a distributed fashion and changes the sampling rates of the CHs based on the variance of the sampled data without damaging significantly the accuracy of the sensed area.**

## I. INTRODUCTION

### A. Background

In the past few years, rapid advances in the area of micro-technology and nano-technology have taken place with implication to all of the scientific research fields. As a result, micro-sensors have been developed for varied needs. Subsequently, this has led to the development of Wireless Sensor Networks (WSNs). WSNs are composed of a variety of nodes, and they include abilities of data sensing and data processing, as well as wireless transmission such as Bluetooth or radio technology. The invention of WSNs has led to the development of various serviceable applications, including control, tracking and monitoring of large areas [1].

The WSN nodes are powered by batteries and are designed to operate without supervision for a relatively long duration of time. However, it is often problematic to replace or recharge the sensor node batteries, due to the fact that most of the deployments are in large fields (e.g. animal control) or inaccessible places (e.g. war zones) [2]. Therefore, energy efficiency is a crucial performance metric in sensor networks, that imminently determines the network lifetime. In order to tackle this array of considerations, some protocols may be used to handle the trade-off between performance metrics (such as network overall postponement) and energy efficiency [3].

Energy may be obtained by utilizing the external environment (for example, by using photovoltaic cells as an energy source). Nevertheless, the behavior of external energy source is usually non-persistent, which makes it unreliable and requires the use of batteries [4]. However, batteries also cause a problem because of their finite amount of stored energy and the frequent need to replace or recharge them. Consequently, a key factor in WSN is to improve overall energy efficiency to extend the network lifetime. Energy conservation influences the design of many WSN-based systems.

Comprehensive studies have examined this issue [5] and suggested energy optimization techniques in order to manage the WSN topology correctly. The most important implications from those studies [5] [6] to our research are:

1. Although in most of the cases there are more only-sensing nodes than sensing-and-communication nodes, the energy consumption of the latter may be crucial to the connectivity of the WSN, and an optimization of the working process of those nodes may be beneficial for the whole network.

2. Although the processing stage is not the most consumable stage in the data routing in WSNs, its energy factor might be larger by more than order of magnitude than the energy factor needed for data transmission [7]. This shows that in order to preserve power in those cases there is a need to take in consideration the processing stage too.

*B. WSN Hierarchical Clustering*

Nodes in multi-hop ad-hoc sensor networks play a dual role as a data originator and data router at the same time. Some of the nodes may not operate properly, which may lead to major topological changes and require the rerouting of some packets and network reconstruction. This further emphasizes the importance of energy efficiency and energy control. Because of this, the focus of many researchers of WSN is to develop protocols and algorithm that consider scalability and energy efficiency by grouping the network nodes into clusters in order to form a hierarchical topology and eliminating any redundant data [8][9]. This process consists of the following steps: Nodes transmit their sensed data to a master node; The master node aggregates the data; The master node performs computations on the data and gets rid of the superfluous data; The master node forwards the new calculated data to the sink node.

A formalized approach to demonstrate this concept assumes that a WSN cluster consists of the following main components: a base station (BS) and a number of sub-clusters. Each of the sub-clusters has a leader (usually referred to as the cluster-head, or CH), as well as other nodes (non-cluster-head nodes, or NCH) that are all within the same transmission range. The transmission range is defined as the maximal distance between a receiver and a sender (in this case, CH and NCH respectively). There is a correlation between the length of the transmission range and the energy consumed in this topology. Numerous advanced transmitter units can modify the transmission energy consumption factor in a dynamic and adaptive manner, in order to enable energy saving in the consumption required to approach nearby receivers. This available low power transmission rates may decrease disturbance, which subsequently increases the network throughput. Yet, one of the limitations of low energy rate can be seen in the connectivity of the network - if the energy transmission rate is low, then each node may have less directly approachable nodes. It is worth mentioning that the techniques available for this kind of energy preservation operate solely on the link layer and do not harm the data processing jobs performed by a node, in contrast to sleep cycle conduct [10].

Another utilization of cluster-based routing includes the formation of energy efficient hierarchical topology in WSNs. If the routing is viewed as a hierarchical topology, it is possible to use CHs, which are the nodes with a higher energy rate, to gather, process and transmit information. The NCHs nodes, which have a lower energy rate, can be used to perform only the sensing. The short-range transmission in clustered WSNs – i.e. to an approximate CH rather than a long-distant BS – may lower energy consumption and eventual premature termination of NCHs. On the other hand, the CHs may need to gather, process and transmit data for the entire cluster, which may affect their performance and reduce their lifetime – particularly in the lack of relay nodes (which ease the burden on the CHs) between CHs and BS [11].

In this paper we present an optimization to the energy consumption of the CHs in a hierarchical WSN topology, using a new modification of the sensing rate, which targets specifically the processing stage. In order to set the ground for this optimization we choose the Hierarchical Control Clustering (HCC) algorithm [12] as our cluster formation and handler layout.

*C. The Hierarchical Control Clustering (HCC) Algorithm*

In order to optimize the energy consumption of a hierarchical WSN topology, we need to determine which clustering scheme will take place along with our optimization method. Unlike most of the published schemes, the goal of Banerjee and Khuller scheme is to form a multi-tier hierarchical clustering using proximity-traversing-based algorithm named Hierarchical Control Clustering [12] (henceforth, HCC). HCC is a distributed multi-hop hierarchical clustering algorithm which also effectively manages to create a multi-level cluster hierarchy. The algorithm works in a distributed fashion, meaning that each node in the WSN can initiate the cluster formation process. The HCC progress in two main sub-processes, when the first is the Tree Discovery process and the second is the Cluster Formation process. The first process is fundamentally a distributed formation of a BFS tree, which is root is the initiator node. The second process is initiated when a sub-tree of a node rises above a predefined size parameter k. Than the node starts cluster formation on its sub-tree. If the sub-tree size is less than 2k, it will form a single cluster for the whole sub-tree. Otherwise, it will form multiple clusters. This two steps process has a time complexity of O(n), even though it has managed to obtain balanced clustering, and to deal with non-stable environments quite effectively.

II. THE ADAPTIVE DISTRIBUTED HIERARCHICAL SENSING (ADHS) ALGORITHM

*A. Energy-Efficient Schemes and Hierarchical Sampling*

Energy-efficient schemes can be mainly divided into two purposes: reducing the energy consumption of all nodes in a WSN, and increasing the WSN lifespan and connectivity. While these two objectives are highly interdependent, they differ because the reduction of overall energy consumption focuses solely on a minimization problem, while the increase of network lifespan demonstrates a min-max problem, as network lifespans may often be fixed or influenced by the network nodes that have the shortest lifespan [3]. The nodes with the shortest lifespan are referred to as *bottleneck nodes* and in the case of clustered WSNs they are often the CHs. In this paper, we adopt the idea of minimizing the overall energy consumption in conjunction with the idea of maximizing the network lifetime. We investigate the amount of time that our optimization can prolong the energy consumption of some $X$ percentage of CHs, and how this affects the specific coverage or connectivity of the region nodes, and the reliability of the sensed data.

Another aspect of energy-efficient schemes is based on the sensors themselves. One of the ways to reduce energy consumption is hierarchical *sampling*. This is a solution based on the actual sensors of a sensing node, and uses the nodes that are equipped with different types of sensors in a differential fashion [13]. Each sensor exhibits different performance abilities in aspects of energy consumption and precision of sensing, and together they provide sufficient functionality for a whole WSN. A trade-off of accuracy and energy consumption can be addressed by using energy-efficient scheme to get coarse-grained data of the general sensing field during idle time, and another energy scheme with higher resolution when some event is detected. Such an approach is known as *triggered sampling* or *load balancing aware sampling* [14].

An example to this technique can be seen in nuclear reactors being split into zones that contain different types of sensors [15], such as the temperature sensors near the reactor core and the radiation contamination sensors that are more distant from the core. The temperature nodes are equipped with heat sensing equipment and they periodically sample the environment, while the radiation contamination sensors include a precise measurement of the radiation outside of the reactor core, and they are in a sleep mode most of the time. When no event is triggered to detect radiation contamination (such as excessive heat inside of the reactor core), the radiation sensors sleep until they are subsequently re-activated in every pre-determined period of time. If an excessive heating was suddenly detected in one of the reactor core temperature sensors, all of the sensors will immediately cross-check the readings of their neighbors and if the readings shows suspicious results, the surrounding, densely deployed radiation contamination sensors will be activated, obtaining fine-grained data about the situation and subsequently indicate and report the damage.

In this paper we adopt the idea of hierarchical sampling and triggered sampling in a different way from the common applications, meaning that although we do not use different types of sensors nor different types of cycles for the whole WSN, we manage to optimize the energy consumption of the CHs by a small modification of the processing and transmitting stages cycles, without damaging the coherence of the sampled data, and with a simple uniform type of sensors which all works at every cycle, as will be presented in the following chapters.

*B. Research Goals*

Unlike other energy-efficient schemes for hierarchical sampling that were mentioned above, our hypothesis is that it is achievable, in terms of prolonging the network lifetime, to adaptively re-modify CHs sensing rates in some specific regions that are triggered significantly less than other regions in a homogenous sensors network in a distributed fashion. Moreover, this can be achieved without damaging significantly the accuracy of the sensed area. In order to achieve that goal, we first use a WSN cluster which is formed by the HCC algorithm, i.e., a hierarchical formation of a WSN. At the beginning, all of the CHs and the NCHs are sampling the area at every cycle. We define this default measurement rate (i.e. processing and transmitting the arrived sensed data) by a variable $C$ which is equal to 1 in each CH and NCH. This means that each node samples its environment in each cycle. Afterwards, in the distributed processing stage of the CHs children sampled data, the CHs check the variance of the gathered data. Then, based on threshold variable $V$ which was set by the user, the CHs will determine whether to increase or decrease their measurement rate $C$. It is important to notice that a change of the measurement rate $C$ is only applicable to the processing and the transmitting of the processed data forward, meaning that although there is a change of $C$, the CHs keeps receiving the data from the NCHs. In case that the variance crosses the variance-threshold $T$ after $C$ was increased, all of the data that has been sensed over the last period of time (in which processing and transmitting was not done) will be processed and sent to the higher CH in the WSN. In this way, the data reliability will slightly compromise (or even not compromise at all) on one hand, and the network CHs will be able to preserve more energy, on the other hand. By doing so, it is possible to minimize or even prevent the premature death of CHs (comparing to the other nodes in the network), which are less burdened with sensing and transmitting actions.

For example [16], imagine that the South-African government decided to hermetically map the movement of the Giraffes in Kruger national park, one of the largest game reserves in Africa, which covers an area of 19,485 square kilometers in the provinces of Limpopo and Mpumalanga in northeastern South Africa. It is known that the Giraffes take part in a long-distance migration, synchronize to overlap with the yearly pattern of rainfall and herbiage sprouting on some several specific plains where they can trace the nutrient fodder [17]. Because of the huge masses of the Giraffe herds and their height, it is impracticable to collar their members with wireless sensors. Consequently, in order to do trace those herds, at the center of each 100 square meters a wireless sensor with animal sound recognition [18] been placed, and initially all of those nodes creates a uniform hierarchical WSN cluster. Now, because the Giraffes tend to move in large-scale herds and not in individual fashion, it means that soon there is going to be some massive parts of the network which will not sense the Giraffes noises for long period of time but will still process and transmit the received non-informative data (Figure 1) - a state which is very expansive, and which will lead to a reduced connectivity of the network for no good.

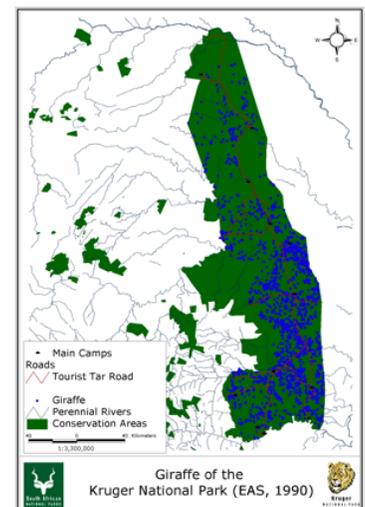

**Figure 1.** Giraffes of the Kruger National Park (EAS, 1990)

*C. Problem Formulation*

We consider a WSN where n sensor nodes have been deployed in a bounded $L \times L$ ($m^2$) square. The location of each sensor $v_i, i \in 0, 1, ..., n-1$, is denoted as $(x_i, y_i)$. We assume a one-hop communication model, and the transmission energy is calculated using the $d^2$ power-loss model. We consider a sensor deployment scenario in a uniform node distribution. The optimization problem is to strategically refine the sampling rate of parts of the entire WSN by calculating the variance of the results sampled during periods of times, and comparing it to a threshold given by the user. By doing so it is possible to reduce the energy consumption of parts of the WSN without significantly damaging the accuracy of the data that should be sampled. Thus, (1) the total energy consumption for the transmission of each unit of data message from all NCHs to CHs and so forth is reduced, (2) the processing energy of the CHs is reduced due to less cycles of sampling, and (3) the total energy consumption of the entire WSN will reduce, and it will be possible to increase the WSN lifetime and connectivity at once.

We consider the following general assumptions in our problem formulation: All sensors are pre-deployed and have constrained energy supply; The network is static, that is, neither the sensors nor the CHs has mobility once deployed; The total number of sensors is known; Each CH forms exactly one cluster, and besides data processing, also performs the same task of environmental sensing and data collection as a regular sensor node; There exists a contention free MAC protocol for wireless communication. We consider the energy consumption for data transmission of each NCH, and for data receiving, processing, and transmission of each CH. Since the energy cost for environment sensing is generally much less than communication and processing tasks, we do not consider sensing energy cost here. Obviously, the total energy consumption depends on the network distribution, the number and location of CHs, and the compression ratio α at CHs.

*D. The ADHS Algorithm*

As previously explained, in order to preserve energy in the non-burdened CHs, a modification of the sensing rate of those specific non-triggered sub-clusters of the whole WSN should be performed. This can prolong the total lifespan of the network. Consequently, we achieve the goal of reducing the overall energy consumption as well as increasing the network lifetime.

| **The Adaptive Distributed Hierarchical Sensing Algorithm (ADHS)** |
|---|
| 0.1  *cycles_counter* = 1 |
| 0.2  *Sensed_data* = *{}* |
| 0.3  $T \in \{0, ..., \infty\}$ |
| 0.4  $L \in \{1, ..., \infty\}$ |
| 0.5  C = $\{c_1, c_2, ..., c_n\}$ = $\{1, 1, ..., 1\}$ |
| 0.6  V = $\{v_1, v_2, ..., v_n\}$ = $\{1, 1, ..., 1\}$ |
| |
| **ADHS**(CH $ch_i$, Thresholds *{T,L}*, variance $v_i$, cycles $c_i$) |
| 1. *if cycles_counter* < $c_i$: |
|   1.1. *sample* the data of $ch_i$ |
|   1.2. *for each* node *NCH* in nodes($ch_i$): |
|     1.2.1. *Sensed_data* = *Sensed_data* ∪ *sample* the data of the *NCH* |
|   1.3. *cycles_counter* = *cycles_counter* + 1 |
| 2. *else:* |
|   2.1. *sample* the data of $ch_i$ |
|   2.2. *for each* node *NCH* in nodes($ch_i$): |
|     2.2.1. *Sensed_data* = *Sensed_data* ∪ *sample* the data of the *NCH* |
|     2.2.2. $v_i$ = *calculate* the variance of the *Sensed_data* |
|   2.3. *if* $v_i \leq T$ and $c_i \leq L$: |
|     2.3.1. *Transmit* last *Sensed_data* |
|     2.3.2. $c_i = c_i + 1$ |
|   2.4. *else* |
|     2.4.1. *process & transmit* the *Sensed_data* |
|     2.4.2. $c_i = 1$. |

Our algorithm, which we refer to as the Adaptive Distributed Hierarchical Sensing algorithm (henceforth, ADHS), makes decisions based on local behavior of clusters, rather than taking into account the behavior of the entire network. In this way we can significantly improve the running time of our algorithm without compromising energy efficiency. The algorithm is executed by all CHs in a distributed fashion. Initially, each CH is provided with a default measurement rate *C*, a threshold variance of the gathered data T and cycles expansion limit *L*, which are set by the user, and a *V* variance of the actual sensed data. In the initial configuration the *C* values are the same (i.e., equal to 1) in all the nodes of the network, and represent a balanced environment. The *V* values are set initially to 1 (i.e., all measurements are not the same). The algorithm can start from any cluster-hierarchy tree, where the simplest configuration is a single node, which functions as a CH. At every cycle, while the variance of the sensed data *V* does not cross the *T* threshold, and *C* does not cross the *L* threshold, we transmit the last calculated sensed data and perform a local refinement in the cluster of that *C* value, increasing it by 1 (another saving in energy consumption may be achieved by giving up on the transmitting stage). Alternatively, when *V* crosses the *T* threshold, or when *C* crosses the *L* threshold, we will make *C* equal to 1 again. Then we will process and transmit the sensed data over that time. This refinement results in a better energy use in each such CH, without jeopardizing the correctness of the sensed results. In other words, we balance clusters of excess energy-use by adaptively changing their sub-clusters sampling rates in accordance to the sensed reality.

III. THE ADHS ALGORITHM ENERGY CONSUMPTION MODEL

*A. WSN Energy Consumption Model*

We consider two different types of energy consumption for data transmission and receiving, respectively: a transmitter consumes energy to run both the radio electronics and the power amplifier, while a receiver only consumes energy to drive the radio electronics. The mobile radio channels on typical sensor nodes are predominantly in the VHF (frequency from 30MHz to 300MHz, wavelength from 1m to 10 m) and UHF (frequency from 300MHz to 3GHz, wavelength from 10cm to 1m), respectively [19][20]. We employ the free space (*fs*) fading channel model for wireless communication that incurs a $d^2$ power loss [19]. In a real communication system, the transmission power could be adjusted by suitably configuring the power amplifier. Therefore, the energy dissipation in transmitting one unit of data message over a directed wireless communication link can be modeled as $E_t(i)$, when $E_t(i) = E_{elec} + E_{amp}(d_{i,j}) = E_{elec} + \epsilon_{fs} \cdot d_{i,j}^2$, where $E_{elec}$ denotes the energy for driving the electronics, which depends on various factors including digital coding, modulation, filtering, and spreading of the signals, for both transmitter electronics and receiver electronics; and $\epsilon_{fs}$ is the coefficient for calculating the amplifier energy $E_{amp}$, which depends on the Euclidean distance $d_{i,j} = \sqrt{(x_i - x_j)^2 + (y_i - y_j)^2}$ between transmitter $v_i$ located at $(x_i, y_i)$ and receiver $v_j$ located at $(x_j, j)$ as well as the acceptance bit-error rate. The energy consumed by a sensor $v_i$ in receiving one unit of data packet is denoted by $E_r(i) = E_{elec}$. Note that the above transmission and receiving energy models assume a contention free MAC protocol, where interferences from simultaneous transmission can be avoided.

A CH, which also collects environment sensing data, receives data messages from NCHs within the cluster and sends all the data to a main CH or BS after performing a certain type of data processing (such as data aggregation and data compression). We use a constant $E_p$ to represent the energy spent in processing each unit of received or sensed data. We assume that the CH performs complete data aggregation, that is, an input of two k-bit messages produces an output of one k-bit message after aggregation. Furthermore, we use a parameter α, $0 < \alpha \leq 1$, to denote the data compression ratio: an input of k bits results in an output of α · k bits after compression.

*B. ADHS Algorithm Energy-Efficiency Proof*

In this paper, we use an analytical approach for calculating the value of the sampling rate of non-loaded parts of the WSN in order to achieve improved total energy consumption of the CHs receiving and processing phases.

The total energy consumption per round, denoted by $E_{Tot}$, is the sum of the energy consumption $E_{NCH}$ of all NCHs for data transmission and the energy consumption $E_{CH}$ of all CHs for data receiving, processing, and transmission in one round, which can be defined as in Formula (1):

$$E_{Tot} = E_{NCH} + E_{CH} \quad (1)$$

The $E_{NCH}$ only includes transmission energy cost $E_T$, when $E_{CH}$ includes the energy cost $E_r$ for receiving, $E_p$ for processing, and $E_t$ for transmission. Each of NCHs transfers one unit of data to its corresponding CH, which performs processing (aggregation and compression) on the received data and its own sensing data, and sends the compressed aggregated result to other CH or BS.

In order to show that our algorithm reduces the total energy consumption of the whole WSN and not harm the connectivity of the network we need to show that (I) the total energy consumption of the data sampling refined zone is actually lower than the previous state, and (II) that the accuracy of the data that has been sampled is close as possible to the real data. Therefore, we need first to formulate the energy consumption of the CHs in our model. Based on the previous knowledge of $E_{Tot}$, for each CH in our model the energy consumption will be:

$$E_{CH} = n_{NCH \to CH} E_r + (n_{NCH \to CH} + 1) E_p + \alpha E_t \quad (2)$$
$$= n_{NCH \to CH} E_{elec} + (n_{NCH \to CH} + 1) E_p + \alpha (E_{elec} + \epsilon_{fs} \cdot d_{CH \to HigherCH}^2)$$

Where $n_{NCH \to CH}$ is the number of NCH that communicate with the CH, and $d_{CH \to HigherCH}$ is the distance between the CH to its higher CH in the hierarchy (the BS is the top node is the hierarchy). Hence, a refinement of the sensing ratio in parts of WSN is always worthwhile if the variance of the data that has been sensed does not cross the variance threshold set by the user, meaning that this data is currently redundant and it is beneficial to change the sensing ratio and avoid near future redundant data. The save of the energy for single round in this situation is done by reducing the energy cost $E_p$ for processing data, while transmission and receiving of data continues as always. A formulation of this condition, based on the formula (2) to $E_{CH}$ will be:

$$E_{CH} = n_{NCH \to CH} E_r + 1 E_p + \alpha E_t \quad (3)$$
$$= n_{NCH \to CH} E_{elec} + 1 E_p + \alpha (E_{elec} + \epsilon_{fs} \cdot d_{CH \to HigherCH}^2)$$

Hence, for a given sample rate $r \in [0,1]$ the average energy consumption per cycle, while $r$ doesn't change, for each single CH will be:

$$E_{CH} = n_{NCH \to CH} E_{elec} + \\ (r \cdot n_{NCH \to CH} + 1) E_p + \\ \alpha(E_{elec} + \epsilon_{fs} \cdot d_{CH \to HigherCH}^2) \quad (4)$$

Based on the known energy consumption parameters [19] $E_{elec} = 5 \cdot 10^{-8} J/bit$, $E_p = 5 \cdot 10^{-9} J/bit/signal$, and $\epsilon_{fs} = 10^{-10} J/bit/m^2$, the energy save in formula (4) comparing to formula (2) is dramatic, especially because of the order of magnitude difference between the different coefficients.

An exemplification of the algorithm and the energy save are demonstrated in figure 2 and figure 3, which presents a 3-level hierarchy WSN grid (1st level = blue CH; 2nd level = yellow CHs; 3rd level = black NCHs) formed by HCC before and after a one round of the ADHS algorithm refinement. Initially (Figure 2), the formed WSN have no data about the surrounding area (the squares with two inverse arrows represents the source of the data), and all of the CHs clocks are initiated to sample the surroundings at every cycle (dotted squares represent this datum).

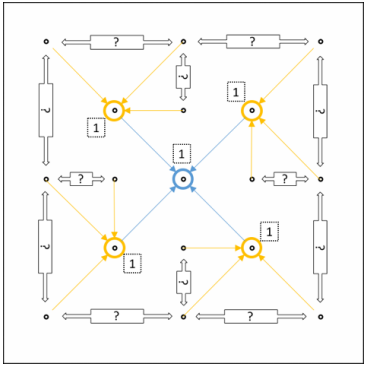

**Figure 2.** A 3-level WSN grid at the initiate state, before ADHS refinement.

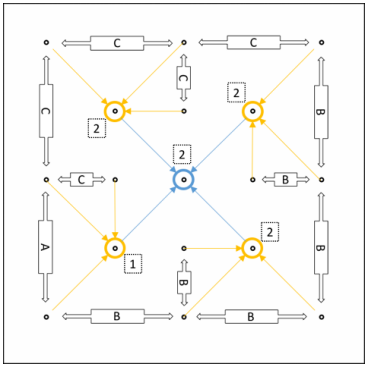

**Figure 3.** A 3-level WSN grid after sampling the surrounding data (A=10, B=20, C=30) and the activation of the ADHS algorithm with threshold T=15.

Afterwards, the WSN starts to sample the surrounding data (Figure 3). In this case we set the surrounding data to three different sets such that set A=10, set B=20 and set C=30, and the user set the variance threshold to 15. It is possible to see in Figure 3 that due to the algorithm execution, 3 CHs in the 2nd level and the main CH in the 1st level changed their sampling ratio. The rest of the CHs which their sampling rate was not changed sampled data with a variance larger than the set threshold (T=15).

As previously explained, in order to examine if the refinement achieved its goals, we need to focus on the amount of $E_r$ and $E_p$ energies that this algorithm saves for the CHs at every cycle. According to formulas (2) and (3) before refinement the total energy of the WSN CHs in the example above was equal to:

$$E_{Total_{CHs}} = 4(3E_r + (3+1)E_p + \alpha E_t) + (4E_r + (4+1)E_p + \alpha E_t) = 16E_r + 21E_p + 5\alpha E_t$$

While after refinement, because of the new sampling rate of some CHs, the WSN total energy of the WSN CHs in the example above is equal to:

$$E_{Total_{CHs}} = 3\left(3E_r + \left(\frac{1}{2} \cdot 3 + 1\right)E_p + \alpha E_t\right) + \\ (3E_r + (3+1)E_p + \alpha E_t) + \left(4E_r + \left(\frac{1}{2} \cdot 4 + 1\right)E_p + \alpha E_t\right) \\ = 16E_r + 14.5E_p + 5\alpha E_t$$

It is possible to see that as long as the variance of the data does not cross the threshold, the sampling rate (the processing and transmitting stages) will keep go down, or until it will reach a barrier which also can be set by the user. In this case there was a ~30% save in the energy consumption of $E_p$ after one execution of the ADHS algorithm.

IV. CONCLUSIONS AND FUTURE WORK

Those results and analysis lead to the conclusion that it would be beneficial to use the ADHS algorithm along with the HCC algorithm in WSNs with differential load for maximization of the network lifetime as well as connectivity. This paper opens a number of prospective directions for future research. One immediate direction is to explore how the ADHS algorithm is reacting along with different WSN hierarchical clustering algorithms, and what exactly does that mean in aspects of cost, complexity, energy-efficiency and connectivity of the network. Another direction is to understand how to optimize other WSN clustering algorithms which are not based on a hierarchical formation of the WSN using the ADHS algorithm.

ACKNOWLEDGMENTS

This work was supported by the Lynn and William Frankel Center for Computer Science, the Open University of Israel's Research Fund, and Israel Science Foundation grant 724/15.